\documentclass[draft]{agujournal2019}
\usepackage{url} 
\usepackage[inline]{trackchanges} 
\usepackage{soul}
\usepackage{amsmath}
\usepackage{lscape}
\usepackage{enumitem}
\draftfalse

\journalname{JGR: Planets}

\begin{document}

%
%


\title{Birth and decline of magma oceans.\\ Part 1: erosion and deposition of crystal layers in evolving magmatic reservoirs.}

%
%




\authors{Cyril Sturtz\affil{1}, Angela Limare\affil{1}, Stephen Tait\affil{1} and \'Edouard Kaminski\affil{1}.}

 \affiliation{1}{Universit\'e de Paris, Institut de Physique du Globe de Paris, CNRS, F-75005, France}





\correspondingauthor{Cyril Sturtz}{sturtz@ipgp.fr}




\begin{keypoints}
	\item We study experimentally the transient behavior of a convective fluid bearing particles that can float/sediment to form deposits.
	\item The erosion process is described by a modified Shields number that compares the convective shear to the beads buoyancy. The deposition rate of crystals scales with the Stokes velocity of beads in absence of convection. 
	\item This theoretical framework yields scaling laws that can be applied to various magmatic systems based on the dimensionless numbers that characterize them.
\end{keypoints}

%
%

%
%


\begin{abstract}
\indent This paper is the first of a two companion papers presenting a theoretical and experimental study of the evolution of crystallizing magma oceans in planetesimals. We aim to understand the behavior of crystals formed in a convective magma and the implications of crystal segregation for the reservoir thermal and structural evolution. In particular, the goal is to constrain the possibility to form and preserve cumulates and/or flotation crusts by sedimentation/flotation of crystals. We first use lab-scale analog experiments to study the stability and the erosion of a floating lid composed of plastics beads over a convective viscous fluid volumetrically heated by microwave absorption. We propose an erosion law that depends only on two dimensionless numbers which govern these phenomena: (i) the Rayleigh-Roberts number, characterizing the strength of convection and (ii) the Shields number, that encompasses the physics of the flow-particle interaction. We further consider the formation of a cumulate at the base of the convective layer due to sedimentation of beads that are denser than the fluid. We find that particles deposition occurs at a velocity that scales with the Stokes velocity, a result consistent with previous experimental studies. The theoretical framework built on these experimental results is applied in a second paper on the evolution of magma oceans in planetesimals and the fate of particles in this convective environment.
\end{abstract}

\section*{Plain Language Summary}
\indent At early time in planetary formation, the heat supplied by $\rm{^{26}Al}$ was large enough to generate massive melting episodes in planetesimals and to produce magma oceans. Following melting and iron core differentiation, the proto-mantle of these 10-100's km radius rocky bodies was a magma ocean that behaved like a convective fluid bearing solid crystals. The goal of this study is to understand the crystal-melt segregation in a magma ocean.  In order to characterize the behavior of crystals in a convective flow we build  a theoretical model that is based on the results of lab-scale experiments. We use a viscous fluid heated by microwave absorption in order to simulate the  convection regime expected in magma oceans. We use plastic beads to simulate crystals in the experiments and study the formation of either a flotation lid or a sedimented basal cumulate. We demonstrate that both deposits form only if the the buoyancy of the fluid balances the  shear imposed by convection, and we predict the transient evolution observed in the experiments. 

%
%

\section{Introduction}
	\indent Crystal segregation in magmatic reservoirs, including magma chambers \cite{Sparks19}, lava lakes \cite{Helz89}, and planetary-scale systems such as magma oceans \cite{Solomatov00}  is a key process in their thermal evolution. Their dynamics can be described as the one of a convective fluid bearing silicate crystals. Because the composition of crystals differs from that of the magma, a density difference always exists between the two phases and induces a buoyancy force whose strength and sign depend on the type of considered crystal. For instance, olivine crystals are usually negatively buoyant and may form a cumulate in magma chambers \cite{Jellinek01}, whereas plagioclase crystals are positively buoyant and may form a flotation mush at the surface \cite{Namur11}. At planetary scale, this segregation process may also occur and heavy crystals can form cumulates of dunites and/or harzburgites at the core-mantle boundary (CMB) by sedimentation \cite{Righter97,Mandler13}, whereas light crystals accumulate and form a crust at the surface, like plagioclases that compose the anorthosite crust of the Moon \cite{Wood70a,Wood70b,Warren85}. \\
	\indent  The fate of a suspension in magma oceans has an overriding influence on its geochemical and geophysical evolution. Studies that deal with the formation of deposits do not usually consider the interaction between the sedimentation and the convection and make the hypothesis of either instantaneous crystal segregation \cite{Hamano13,Maurice20} or stable suspension of crystals in the liquid \cite{ElkinsTanton11,Bryson19}. However, the dynamics of such deposition/flotation processes in a convective system are not well constrained. \citeA{Suckale12a,Suckale12b} proposed a model that takes crystal-fluid interaction into account by developing a numerical study describing the suspension and applied their method to constrain the formation of the lunar crust. However, as this model is based on the direct numerical resolution of the conservation of mass, momentum and energy, it does not provide a physical criterion that can be used in independent studies to determine the possibility of forming a deposit or not. Hence, we aim to provide a general criterion that predicts the ability of forming a crust and a cumulate.\\
	\indent A stability criterion for a suspension has been proposed by \citeA{Solomatov93b}, but gives only the maximal concentration of particles that can be sustained by convection at steady state. \citeA{Sturtz21}  revisited this criteria by adapting the physical framework describing the stability of grains in a bed-load \cite{Shields36,Charru04} to the stability of a bead layer sheared by thermal convection in an internally heated system. They used experiments to establish that a modified Shields number xan be used to determine if at steady state a particles layer is stable or not, depending on the Rayleigh-Roberts number characterizing the convective vigor and on particle properties. However, \citeA{Sturtz21} did not describe the dynamics of erosion and deposition, and in particular the characteristic timescales that are involved. The present paper proposes this improvement. \\
 	\indent In the following, we model the dynamics of particle erosion and deposition in order to study the time evolution of a suspension and associated particle layers. Based on an experimental study, we analyse the thermal evolution of a convective system that is insulated by an erodible lid of particles. We show how the framework developed by \citeA{Sturtz21} at steady state can be adapted to describe the thermal evolution of the system and the lid thickness dynamics. These results will be used in a companion paper to constrain the evolution of magma oceans and the formation of layered mantles in planetesimals. 
 
%
 
\section{Experimental approach}

%
%

	\subsection{Experimental set-up}

\begin{figure}
	\centering
	\includegraphics[width=0.6\textwidth]{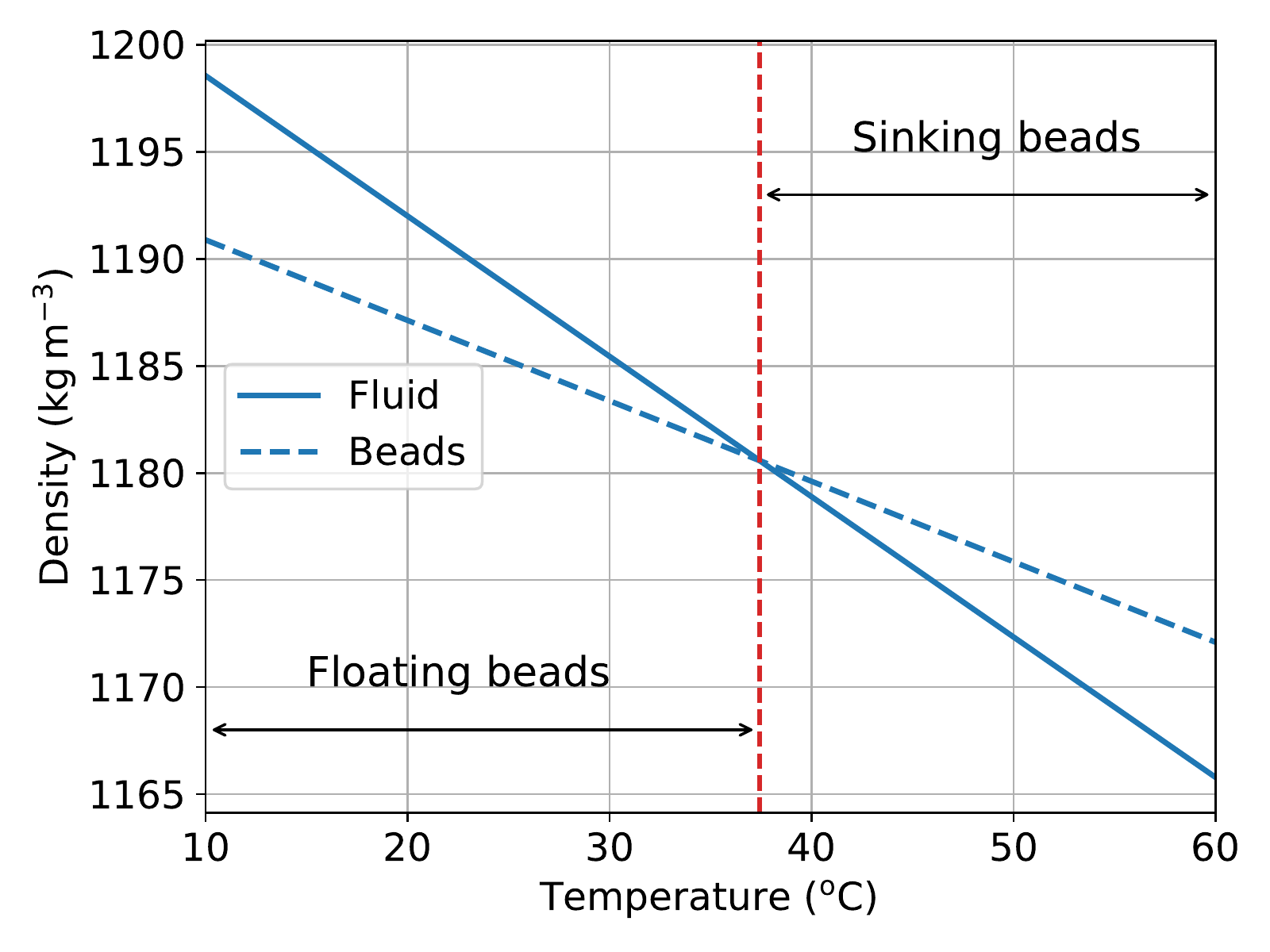}
	\caption{Variation of the respective density of the beads and of the experimental fluid as a function of the temperature, taken from \citeA{Sturtz21}. The thermal expansion coefficient is $\alpha_p=3.17\, 10^{-4}\ \rm{K^{-1}}$ and $\alpha_f=5.5\, 10^{-4}\ \rm{K^{-1}}$ for the beads and the fluid respectively. Beads are positively buoyant for $T<37.4\rm{^oC}$ and negatively buoyant for $T>37.4\rm{^oC}$.}
	\label{fig:density}

\end{figure}

	\indent We study the thermal and mechanical evolution of a convecting viscous fluid under an erodible lid using the method described in \citeA{Sturtz21}. A $300\times300\times50\, \rm{mm^{3}}$ tank is placed in a modified microwave oven \cite{Surducan14}. The top surface of the tank is composed of an aluminium plate whose temperature is fixed and monitored. The lateral and basal walls of the tank are made of plastic (poly-methyl metacrylate, PMMA), which insulate thermally the interior of the reservoir. The fluid inside the tank is heated volumetrically by microwaves absorption, that generates convection at high Rayleigh numbers \cite{Limare13}.\\
	\indent To image the convective fluid, a laser sheet scans the tank, and the signal is monitored by two CCD cameras equipped with filters sensitive to different spectral ranges. The temperature field is measured using a two-dye laser induced fluorescence method, and the velocity field is obtained by particle image velocimetry (PIV). The spatial resolution for each field is  0.2 and 0.8 mm respectively. For a complete description of the experimental methods, the readers can refer to \citeA{Fourel17}.\\
	\indent  The particles used are PMMA beads, and the fluid is a mixture of 44 wt$\%$ glycerol and $56$ wt$\%$ ethylene-glycol. The beads and the fluid have different thermal expansion coefficients (see Table \ref{tab:ParamsExp}, \ref{App:Exp}). As shown in Figure \ref{fig:density}, at low temperatures beads are lighter than the fluid and can float. Above the inversion temperature $T_{\rm{inv}}=37.4\rm{^oC}$, beads become negatively buoyant and can sink. We kept the surface temperature below $37.4\rm{^{o}C}$ so that a floating lid can form at the surface of the convective layer. Meanwhile, the bulk is heated by microwaves and its temperature can eventually exceed  the inversion temperature; in these conditions a basal cumulate can form. The experimental system can thus be used to study the formation and stability of both a flotation crust and/or a basal cumulate.

%
%

\begin{landscape}	

 \begin{table}[h]
   \centering
   \begin{tabular}{c  cccccccccc  }
      \hline
      Name 				& Beads radius	($\mu$m)		&  $\delta_0$ (mm)	& $T_s$ ($^o$C)	& $T_{bulk}$ ($^o$C)		&	 $Ra_H$ ($10^7$)  	& Erosion?	& Cumulate? &    Heat bump?	\\
      \hline

      IHB04         		 	& 	$290$				&		5.3		&	21.8			&	44.5					&		4.3			& Partial		&	yes		& 	no		\\
      IHB05		  	 	& 	$290$				&		5.3		&	21.8			&	47.7					&		8.9			& Partial		&	yes		&	no		\\
      IHB07 	     	 	&	 $290$				&		1.9		&	22.3			&	35.9					&		2.8			&Partial		&	no		&	no		\\
      IHB08 	     	 	&	 $290$				&		1.9		&	14			&	32.8					&		2.4			&Partial		&	no		& 	no		\\
      IHB09 	     	 	&	 $290$				&		1.9		&	34.1			&	43.4					&		4.1			&Partial		&	yes		&	no		\\
      IHB11		    	 	&	 $290$				&		4.7		&	22.8			&	43.9					&		4.2			&Partial		&	yes 		& 	no		\\
      IHB12		    	 	&	 $290$				&		4.7		&	22.8			&	35.3					&		1.4			&Partial		&	no		& 	no		\\
      IHB13 	    	 	&	 $290$				&		4.7		&	23			&	48.3					&		8.0			&Partial		&      yes		& 	no		\\
      IHB14		    	 	&	 $290$				&		4.7		&	8.9			&	42.0					&		5.9			&Partial		&	no		&	yes		\\
      IHB16				&	 $290$ 				&		3.8		&	22.3			&	53.7					&		14.4			&Partial		&	yes		&	no 		\\
      IHB17				&	 $290$				&		3.8		&	22.2			&	38.7					&		2.8			&Partial		&	no		&	yes		\\
      IHB19				&	 $145$ 				&		4.7		&	22.9			&	42.1					&		6.8			&Total		&	no		&	yes		\\
      IHB20				&	 $145$ 				&		4.7		&	29.2			&	45.7					&		8.1			&Total		&	no		&	yes		\\
      IHB21				&	 $145$ 				&		4.7		&	23.1			&	30.6					&		1.5			&Total		&	no		&	yes		\\
      IHB22				&	 $145$ 				&		4.7		&	30.5			&	45.3					&		10			&Total		&	yes		&	yes		\\
      IHB23				&	 $145$ 				&		4.7		&	23.2			&	33.1					&		2.6			&Total		&	no		&	yes		\\
      IHB24				&	 $145$ 				&		4.7		&	23.1			&	26.3					&		0.27			&Partial		&	no		&	yes		\\
      IHB25				&	 $145$ 				&		4.7		&	23.2			&	28.5					&		0.67			&Total		&	no		&	yes		\\
      IHB26				&	 $145$ 				&		4.7		&	33.2			&	36.3					&		1.0			&Total		&	no		&	yes		\\
      IHB27				&	 $145$ 				&		4.7		&	34.4			&	43.7					&		4.4			&Total		&	no		&	yes		\\
      IHB33				&	 $145$ 				&		6.0		&	22.7			&	36.8					&		3.1			&Partial		&	no		&	yes		\\
	\hline
   \end{tabular}
   \caption{Experimental conditions: the beads average radius $r$ (note that two families of beads are investigated), the initial floating bed thickness $\delta_{0}$, the imposed surface temperature $T_{s}$, the mean bulk temperature at steady state $T_{\mathrm{bulk}}$, the Rayleigh-Roberts number calculated at steady state $Ra_{H}$. The next two columns inform about the erosion of the floating lid (whether partial or total), and the formation of a cumulate at the bottom of the tank. The last column deals with the shape of the thermal evolution and designates whether the system experienced a heat bump or not.}
   \label{tab:Exp}
\end{table}		
\end{landscape}

	\subsection{Typical experiments - qualitative behavior of the floating lid}
\begin{figure}
	\centering
	\includegraphics[width=\textwidth]{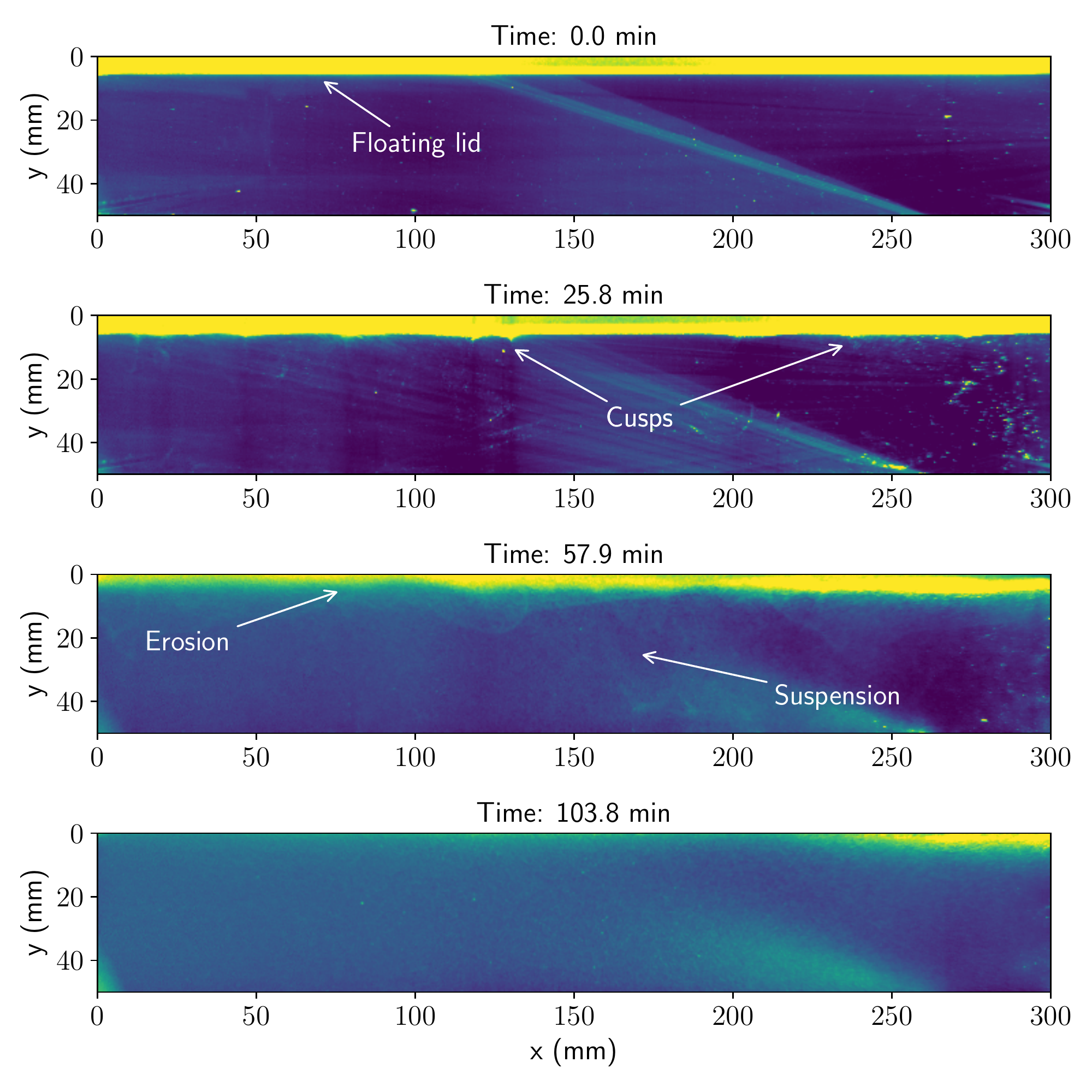}
	\caption{Snapshots from experiment IHB33 ($Ra_{H}=3.10^{6}$, $T_{s}=23\rm{^{o}C}$) showing the destabilization of the floating lid. Dune-shape cusps form once beads have begun to move. Erosion is then triggered and the lid begins to thin. The complete evolution of the lid along the experiment is shown in the movie available in supplementary materials.}
	\label{fig:snaps}
\end{figure}
	\indent During experiments (summarized in Table \ref{tab:Exp}), we record the evolution of the floating lid thickness in order to study the mechanism of destabilization. Snapshots of the lid are displayed in Figure \ref{fig:snaps} and the complete movie is available in supplementary materials. Convection occurs quasi instantaneously in all experiments. A bed-load at the base of the floating lid forms after a delay of a few to several tens of minutes. At that stage, particles move at the surface of the bed and form dunes. This mechanism facilitates beads re-suspension, following the mechanism described by \citeA{Solomatov93a}. Once erosion has been triggered, the lid thins until it reaches an equilibrium thickness at steady state. The final thickness is zero in case of total erosion. We observe that the destabilization of the lid obeys an erosion mechanism, rather than a mechanism involving oscillation of domes and basins as it can be the case for convection in stratified fluids \cite{Jaupart07,Limare19}. This observation prompts to the modeling of the lid evolution by an erosion mechanism such as the one described in the next section.

%
	
\section{Thermal convection under an erodible lid}

%

	\subsection{Convection driven by internal heating}
	
\begin{figure}
	\centering
	\includegraphics[width=0.6\textwidth]{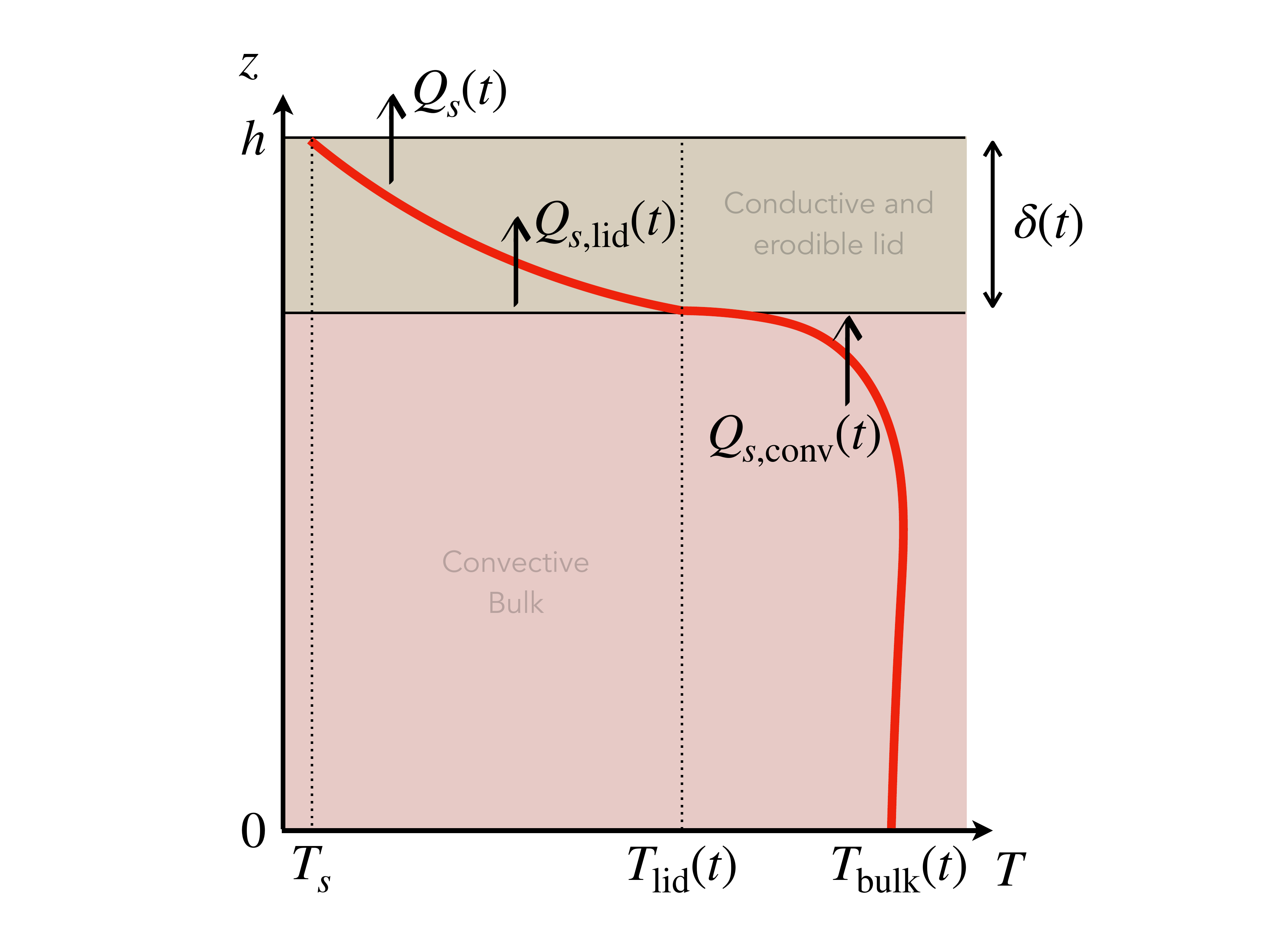}
	\caption{Schematic view of the evolving system. The red solid line represents the horizontally averaged temperature profile, $T_{\rm{lid}}$ is the basal temperature of the lid, $T_s$ is the surface temperature, $Q_s$ is the surface heat flux, $Q_{s,\rm{lid}}$ is the heat flux that enters the conductive lid, $Q_{s,\rm{conv}}$ is the heat flux out of the convective bulk and $\delta$ the thickness of the floating lid.}
	\label{fig:schema}
\end{figure}

	\indent In a fluid internally heated, the temperature scale $\Delta T_H$ is \cite{Roberts67}:
	\begin{linenomath*}
	\begin{equation}
		\Delta T_H=\frac{Hh^2}{\lambda_{\rm{f}}}, \label{eq:DTH}
	\end{equation}
	\end{linenomath*}
	with $h$  the reservoir thickness, $H$  the rate of internal heating and $\lambda_{\rm{f}}$ the fluid thermal conductivity (table \ref{tab:ParamsExp}). Convection is fully characterized by two dimensionless numbers: the Rayleigh-Roberts number $Ra_H$ and the Prandtl number $Pr$ defined as follows:
	\begin{linenomath*}
		\begin{eqnarray}
			Ra_H&=&\frac{\alpha_{\rm{f}}\rho_{0,\rm{f}} g Hh^5}{\eta_{\rm{f}}\kappa_{\rm{f}}\lambda_{\rm{f}}}, \label{eq:RaH}\\
			Pr&=&\frac{\eta_{\rm{f}}}{\rho_{0,\rm{f}}\kappa_{\rm{f}}},
		\end{eqnarray}
	\end{linenomath*}
	where $\alpha_{\rm{f}}$ is the fluid thermal expansion coefficient, $\rho_{0,\rm{f}}$ its density at a reference temperature, $g$ the acceleration of gravity, $\eta_{\rm{f}}$ the fluid dynamical viscosity, and $\kappa_{\rm{f}}$ the fluid thermal diffusivity. In an internally heated convective system, there is a single thermal boundary layer (TBL) at the top boundary of the convective layer. In the experiments performed in this study, $Ra_H$ is high enough for the average bulk temperature to be quasi-isothermal, except in the upper thermal boundary layer (Figure \ref{fig:schema}).  In the high-$Pr$ limit, its thickness of the TBL $\delta_{TBL}$ and the drop of temperature across it $\Delta  T_{TBL}$ at steady state scale with $Ra_H$ as follows \cite{Limare15}:
	\begin{linenomath*}
		\begin{eqnarray}
			\delta_{TBL}&=&C_{\delta}\, h\, Ra_H^{-1/4},  \label{eq:dTBL}\\
			\Delta T_{TBL}&=&C_T\, \Delta T_H\, Ra_H^{-1/4}, \label{eq:DTBL}
		\end{eqnarray}
	\end{linenomath*}
	with $C_T=3.41$ \cite{Vilella18} and $C_{\delta}=7.36$ \cite{Limare15} constants that only depend on the mechanical boundary condition at the top. \\
	\indent If a lid of thickness $\delta(t)$ is present at the top boundary, convection occurs in a the fluid layer of thickness $h-\delta(t)$ and the conservation of energy is written \cite{Kerr90a,Kerr90b}:
	\begin{linenomath*}
		\begin{equation}
			\rho_{\rm{f}} c_{p,\rm{f}} \frac{\partial T_{\rm{bulk}}}{\partial t} = H-\frac{Q_{s,\rm{conv}}}{h-\delta}, \label{eq:ConsNRJ}
		\end{equation}
	\end{linenomath*}
	with $T_{\rm{bulk}}$ the bulk volume averaged temperature, $c_{p,\rm{f}}$ is the fluid specific heat, and $Q_{s,\rm{conv}}$ is the heat flux out of the convective layer. \\
	\indent At steady state, $Q_{s,\rm{conv}}$ equals $(h-\delta)H$, a relationship that does not hold if the temperature of the system or the thickness of the lid are not constant. To determine a general expression for $Q_{s,\rm{conv}}$, we first introduce the modified rate of internal heating $H^*$, including both internal heating and secular cooling effects \cite{Vilella17,Kaminski20,Limare21}:
	\begin{linenomath*}
		\begin{equation}
			H^*=H-\rho_f c_{p,f} \frac{\partial T_{\rm{bulk}}}{\partial t},
		\end{equation}
	\end{linenomath*}
	and we furthermore consider that the temperature difference between the bulk and the base of the lid,  $T_{\rm{bulk}}-T_{\rm{lid}}$, scales as the temperature drop through the TBL given by (\ref{eq:DTBL}):
	\begin{linenomath*}
		\begin{equation}
			T_{\rm{bulk}} -T_{\rm{lid}}=C_T^* \Delta T_H^*\, (Ra_H^*)^{-1/4}, \label{eq:Tlid}
		\end{equation}
	\end{linenomath*}
	with $C_T^*$ a constant that can depend on the mechanical boundary conditions, and $Ra_H^*$ the modified Rayleigh-Roberts number obtained by substituting $H$ by $H^*$ in (\ref{eq:RaH}). After some algebra, we get the following equation for the time-dependent evolution of the bulk fluid temperature:
	\begin{linenomath*}
	\begin{equation}
		\rho_{\rm{f}} c_{p,\rm{f}} \frac{\partial T_{\rm{bulk}} }{\partial t}= H-\frac{1}{(C_T^*)^{4/3}}\frac{\lambda_{\rm{f}}}{h-\delta}\left(\frac{\alpha_{\rm{f}} \rho_{0,\rm{f}} g}{\kappa_{\rm{f}} \eta_{\rm{f}}}\right)^{1/3}\, ( T_{\rm{bulk}} -T_{\rm{lid}})^{4/3}. \label{eq:FLUX}
	\end{equation}
	\end{linenomath*}
	Comparing (\ref{eq:ConsNRJ}) and (\ref{eq:FLUX}), we further obtain a the scaling law for $Q_{s,\rm{conv}}$:
		\begin{linenomath*}
		\begin{equation}
			Q_{s,\rm{conv}}=\frac{\lambda_{\rm{f}}}{(C_T^*)^{4/3}}\, \left(\frac{\alpha_{\rm{f}} \rho_{0,\rm{f}} g}{\kappa_{\rm{f}} \eta_{\rm{f}}}\right)^{1/3}\, ( T_{\rm{bulk}} -T_{\rm{lid}})^{4/3}, \label{eq:ScalQs}
		\end{equation}
	\end{linenomath*}
	with $C_T^*=3.59\pm0.15$ according to \citeA{Limare21}.

%
	
	\subsection{Evolution of the erodible lid}

%
%
\indent We now describe  the evolution of the floating lid, whose thickness and temperature are controlled by the coupling between the lid and the convective fluid in two ways: (i) thermally, by the heat flux that is extracted from the convective bulk and that imposes a thermal boundary condition at the base of the lid, (ii) mechanically, by the erosion process occurring at the interface.
	
		\subsubsection{Thermal state of the floating lid}
	\indent To determine the thermal state of the lid, we solve the Cauchy problem set as follows. The temperature field in the lid $T_b(z,t)$ follows the heat equation for conduction without internal heating source, noting that beads do not absorb microwaves: 
\begin{linenomath*}
	\begin{equation}
		\frac{\partial T_b(z,t)}{\partial t} = \kappa_b\, \frac{\partial^2 T_b(z,t) }{\partial z^2}, \label{eq:TS_b1}
	\end{equation}
\end{linenomath*}
	where $\kappa_b$ is the lid thermal diffusivity. This equation is subject to the initial and surface conditions:
\begin{linenomath*}
	\begin{eqnarray}	
		T_b(z,t=0)&=&T_s. \label{eq:TS_b4}\\
		T_b(z=0,t)&=&T_s, \label{eq:TS_b2}
	\end{eqnarray}
\end{linenomath*}
	with $T_s$ the surface temperature. The basal condition is:
\begin{linenomath*}
	\begin{equation}
		\lambda_{\rm{b}}\, \frac{\partial T_b(z,t=0)}{\partial z}=Q_{s,\rm{lid}} \label{eq:TS_b3}
	\end{equation}
\end{linenomath*}
	with $\lambda_b$ the lid thermal conductivity, and $Q_{s,\rm{lid}}$ the heat flux at the base of the lid. The global transient thermal state of the system is shown in Figure \ref{fig:schema}. One should note that at this stage we do not write a priori that the basal heat flux is equal to the heat flux out of the convective bulk $Q_{s,\rm{conv}}$, as the thermal boundary layer at the base of the lid and the top of the convective layer has to be taken into account. Following \citeA{Kerr90a,Kerr90b}, heat conservation at the moving interface yields:
\begin{linenomath*}
	\begin{equation}
		Q_{s,\rm{lid}}(t)=\rho c_p (T_{\rm{bulk}}-T_{\rm{lid}})\, \frac{\rm{d}\delta}{\rm{d} t}+Q_{s,\rm{conv}}(t). \label{eq:HeatContinuity}
	\end{equation}
	\end{linenomath*}
	The left side term stands for the variation of internal energy of the small moving thermal boundary layer due to change of temperature. In our experiments, the first term in the righthand side is negligible compared to the flux out of the convective bulk (see \ref{App:continuity}) and we take $Q_{s,\rm{lid}}(t)=Q_{s,\rm{conv}}(t)$.
	
%
	
	\subsubsection{Erosion/deposition model}
	
	\indent To describe and quantify the erosional thinning of the lid in our experiments, we use the formalism of \citeA{Shields36}. Within this framework, we consider that particles are subject to  two forces: (i) the convective shear stress that tends to entrain particles, (ii) the friction forces, proportional to the bead's buoyancy, that tends to maintain the particles at the surface. The ratio between these two forces defines the Shields number at the base of the lid:
	\begin{linenomath*}
		\begin{equation}
			\zeta_{\rm{lid}}=\frac{\eta_{\rm{f}} \dot \gamma}{\Delta \rho(T_{\rm{lid}})gr}, \label{eq:shields0}
		\end{equation}
	\end{linenomath*}
	where $\dot \gamma$ is the strain rate due to convection, and $r$ is the bead radius. Erosion starts when the Shields number is larger than a critical value $\zeta_c$.\\
	\indent \citeA{Sturtz21} showed experimentally that the volume average root mean square (RMS) strain rate scales with $\kappa/h^2\, Ra_H^{*,3/8}$ so that the Shields number can be expressed as a function of the characteristics of convection:
	\begin{linenomath*}
		\begin{equation}
			\zeta_{\rm{lid}}=\frac{\eta_{\rm{f}}\kappa_{\rm{f}}}{\Delta \rho(T_{\rm{lid}})grh^2}\, Ra_H^{*,3/8}, \label{eq:shields}
		\end{equation}
	\end{linenomath*}	
and further obtained experimentally that for internally heated convective system $\zeta_c=0.29\pm0.17$.\\
	\indent To obtain an evolution equation for the lid thickness $-$ which will be used later to describe the full evolution of the system $-$ we carry out the mass balance of particles at the base of the lid following the erosion-deposition model \cite{Charru04,Lajeunesse10}. The evolution of the lid thickness is given by the balance between the flux of particles that settle at the lid per unit area $\phi_d$ and the flux of eroded particles per unit area $\phi_e$, i.e.:
	\begin{linenomath*}
		\begin{equation}
			\frac{\rm{d}\delta}{\rm{d}t}=\phi_d-\phi_e. \label{eq:EroDep}
		\end{equation}
	\end{linenomath*}
	 As the lid interface is located in the thermal boundary layer where the convective movements are cold downwellings, we assume that if particles are eroded, they are not able to re-settle as they are entrained by cold downwellings. Hence we take $\phi_d\approx0$. Besides, the flux of eroded particles $\phi_e$ is given by the product of the volume flux of fluid drained by the downwellings $\phi_v$ and the volume fraction of particles in the TBL $c_{\rm{TBL}}$:
	 \begin{linenomath*}
	\begin{equation}
		\phi_e \sim \phi_v\, c_{\rm{TBL}}.
	\end{equation}
\end{linenomath*}
The volume that is drained by the downwellings from the TBL depends on their characteristics. If $N_i$ is the number of downwellings per unit area, $A_i$ their cross section and $W_i$ their vertical velocity, then the volume flux of fluid that is drained per unit area is given by $\phi_v\sim N_i A_i W_i$. Using scaling laws proposed by \citeA{Vilella18} : $N_i\sim h^{-2}\, Ra_H^{1/4}$, $A_i\sim h^2\, Ra_H^{-3/8}$ and $W_i\sim \kappa/h\, Ra_H^{3/8}$, we get:  
\begin{linenomath*}
	\begin{equation}
		\phi_v \sim \frac{\kappa}{h}\, Ra_H^{*,1/4}. \label{eq:PHIV}
	\end{equation}
\end{linenomath*}
Besides, the volume fraction of particles in the TBL $c_{\rm{TBL}}$ is given by:
	\begin{linenomath*}
		\begin{equation}
			c_{TBL}\sim \frac{v_1 \, n}{\delta_{TBL}}, \label{eq:CTBL}
		\end{equation}
	\end{linenomath*}
	with $n$ the volume of eroded particles per unit area, and $v_1$ the volume of one bead. Following \citeA{Lajeunesse10}, we assume that the number of particles per unit area $n$ that are eroded scales with the difference between $\zeta_{\rm{lid}}$ and the critical value:
	\begin{linenomath*}
		\begin{equation}
			n\sim \frac{1}{r^2}\, (\zeta_{\rm{lid}}-\zeta_c). \label{eq:n}
		\end{equation}
	\end{linenomath*}
Combining (\ref{eq:EroDep})-(\ref{eq:n}), we get the following lid thickness evolution:
	\begin{linenomath*}
		\begin{equation}
			\frac{\rm{d} \delta}{\rm{d}t} =-c_e\, \frac{\kappa r}{h^2}\, Ra_H^{*,1/2}\, (\zeta_{\rm{lid}}-\zeta_c), \label{eq:LidEvolution}
		\end{equation}
	\end{linenomath*}
with $c_e$ a constant to be determined experimentally. 

%
%
	\begin{figure}
		\centering
		\includegraphics[width=0.8\textwidth]{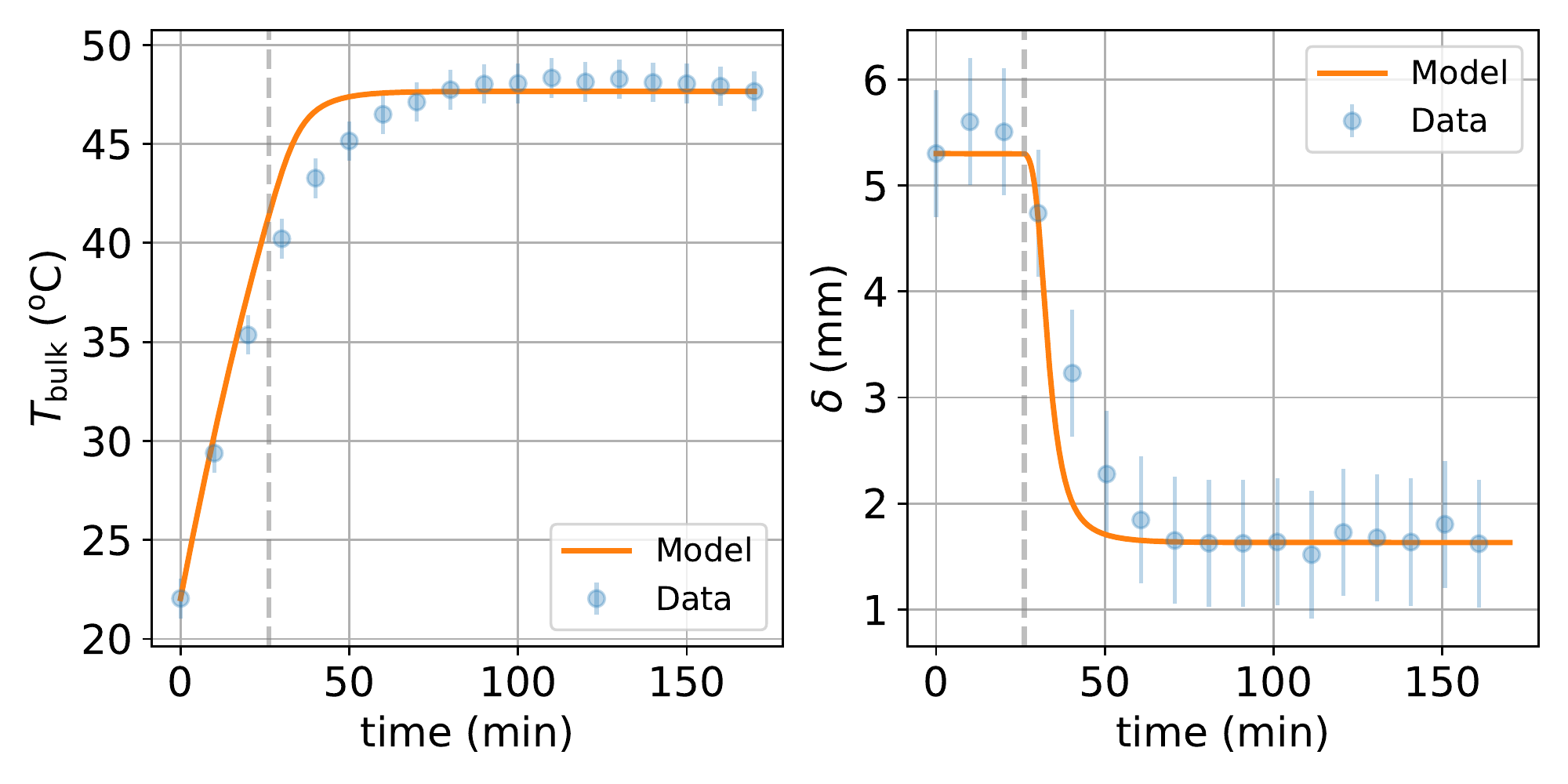}
		\includegraphics[width=0.8\textwidth]{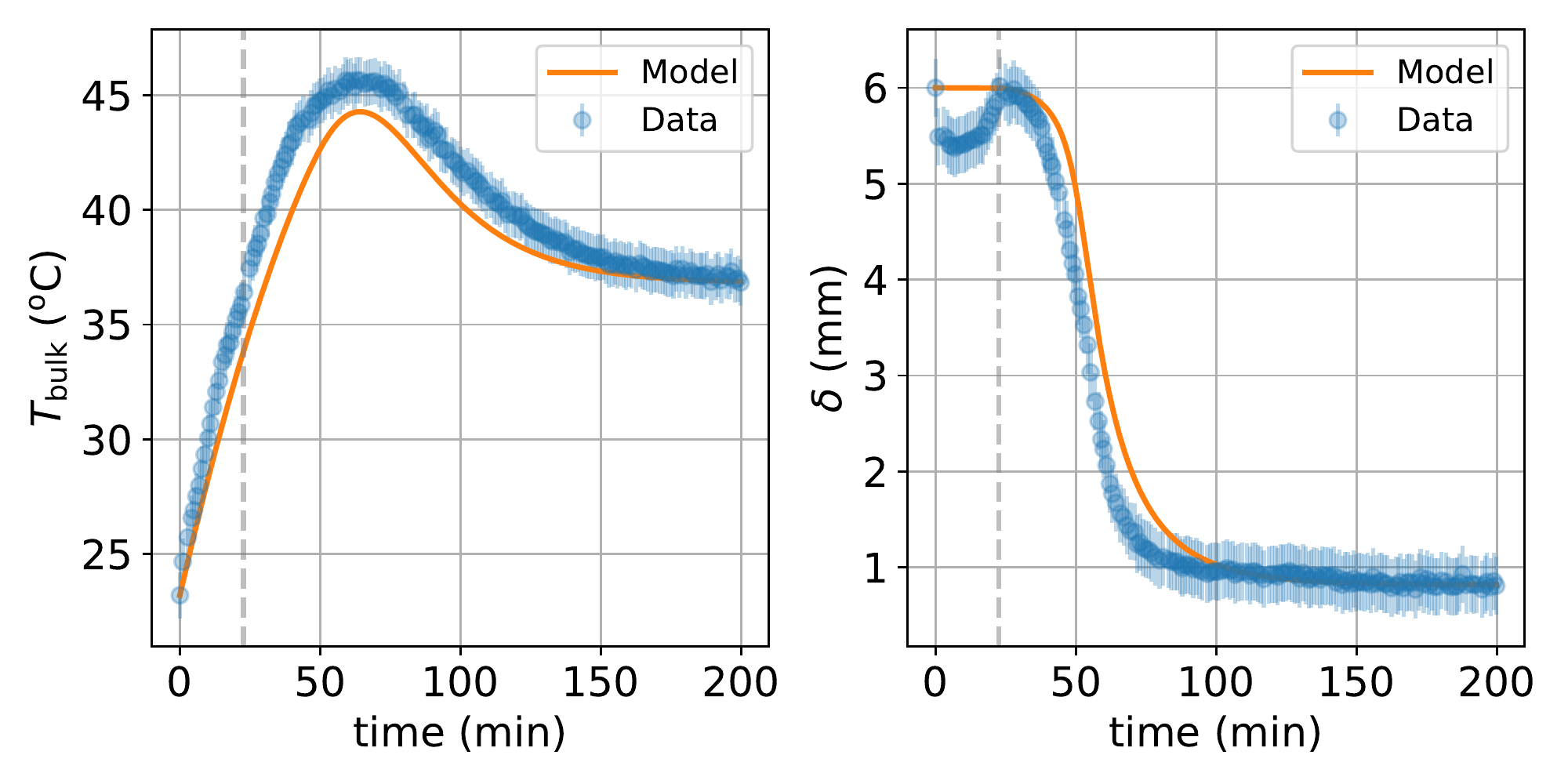}
		\caption{Comparison of experimental data and model prediction for two representative experiments (IHB05 top, and IHB33 bottom). The best fit parameter $c_e$ is $0.1$ and $1$ for IHB05 and IHB33 respectively. }
		\label{fig:NumExp}
	\end{figure}

\section{Comparison between model predictions and experimental data}

%
%

	\subsection{Bulk temperature record}
	\indent  Figure \ref{fig:NumExp} shows the two families of thermal evolution that are observed in experiments. The first evolution is described by a monotonic increase of the temperature towards a steady state thermal plateau reached the steady state. The second one is characterized by an episode of transient thermal maximum followed by a monotonic decrease towards a steady state. To confirm the ability of the model to reproduce these two kinds of evolution, we solve numerically the evolution equations using as input parameters: the surface temperature $T_s$, the rate of internal heating $H$, the initial thickness of the floating lid $\delta_0$, and the final thickness $\delta_{\rm{th}}$. Note that the final thickness is not a direct input parameter but rather deduced from the steady state, assuming conductive heat transfer in the floating lid (see \citeA{Sturtz21}):
	\begin{linenomath*}
		\begin{eqnarray}
			Hh&=&\lambda_b\, \frac{T_{\rm{lid}}-T_s}{\delta_{\rm{th}}},\\
			\frac{\delta_{\rm{th}}}{h}&=&\frac{\lambda_b}{\lambda_{f}}\ \left(\frac{T_{\mathrm{bulk}}-T_{\mathrm{s}}}{\Delta T_{H}}-C_{T}Ra_{H}^{-1/4}\right), 
		\end{eqnarray}
	\end{linenomath*}
	where $T_{\rm{lid}}$ is given by (\ref{eq:Tlid}). We solve (\ref{eq:ConsNRJ}) and (\ref{eq:LidEvolution}) using an explicit finite differences scheme  and we solve (\ref{eq:TS_b1})-(\ref{eq:TS_b3}) using an implicit finite difference scheme. At each step, we calculate the Shields number using the scaling law (\ref{eq:shields}) to obtain the erosion rate. More details on the numerical scheme are given in \ref{App:NS}.\\
	\indent By construction our model gives the coupled evolution of $T_{\rm{lid}}$ and $\delta$. The comparison between the evolution of $\delta$ during an experiment and the model prediction thus provides an independent test of the model's consistency. In order to measure the lid thickness over time, we analyse the light scattered by the lid. As illustrated in Figure \ref{fig:snaps}, when the beads are compacted and settled, they scatter more light than when they are in suspension. Thus we use the number of pixels of high intensity at the top of the frames as a proxy of the total intensity scattered by the lid. The thicker the bed, the higher the intensity scattered by the floating lid and the higher the number of pixels counted. To calibrate the link between the number of pixels of high intensity $\mathcal{I}(t)$ and the thickness of the bed $\delta(t)$, we measured the number of pixels of high intensity at the beginning of the experiment $\mathcal{I}_0$ and at the end $\mathcal{I}_f$. Assuming a linear dependence between the scattered intensity and the thickness of the bed, the thickness of the floating lid is obtained through:
	\begin{linenomath*}
	\begin{equation}
		\delta(t)=\delta_0+(\delta_{th}-\delta_0).\frac{\mathcal{I}_0-\mathcal{I}(t)}{\mathcal{I}_0-\mathcal{I}_f}
	\end{equation}
	\end{linenomath*}	
	We then compare the numerical simulations of the floating lid thickness evolution with the experimental results, as illustrated in Figure \ref{fig:NumExp}, right panels. We note that contrary to the thermal case, there is only one kind of evolution of the lid thickness: a transient period during which the lid remains constant, followed by a delayed monotonic erosion of the lid until it reaches the thickness at steady state.\\
	\indent The good agreement between experimental data and model prediction shown in Figure \ref{fig:NumExp} is verified in all experiments if $c_e$ is allowed to vary between $0.06$ and $2$, leading to an average value $c_e=1.0\pm0.8$, which confirms that our model encompasses the main physical characteristics of the system. Data do not allow us to identify a clear second order dependence of $c_e$, for instance with respect to $Ra_H$. We thus conclude that the large variability of $c_e$ mainly reflects the sensitivity of the model on the determination of $T_{\rm{lid}}$ and $\delta_{\rm{th}}$. We further note that here we describe the erosion of the bed thanks to a simplified 1D model, whereas erosion induces a topography at the lid interface. Hence the variation of $c_e$ could reflect this local bias in the estimate of $\delta$ for example. Moreover, other parameters usually affects the erosion mechanism, such as the packing of the bed \cite{Agudo12} and we further measure a dispersion in beads radius (see \citeA{Sturtz21}, supplementary materials) that can also induce a variability in the determination of the Shields number $\zeta_{\rm{lid}}$.\\
	\indent In both cases illustrated in Figure \ref{fig:NumExp}, we note a small increase of the lid thickness before the erosion begins. This transient thickening can be explained by several factors. At the onset of convection, under the action of shear stress, if the bed is compacted enough, beads are not free to move one around the others. Consequently, the first stage of evolution is a dilatation (i.e. lid thickneing) that enables the motion of beads \cite{Reynolds85,Behringer18}. Another explanation can come from the mechanism of entrainment. As described by \citeA{Solomatov93a}, before being entrained, beads are moving through the lid interface to form dune-shape cusps. This can lead beads to reach the wall and accumulate in the upper corners of the tank. The formation of this small meniscus of beads at the beginning of our experiment could explain the apparent thickening of the lid which might be an artefact in this case.

%
%

\section{Deposition dynamics}
\begin{figure}
	\centering
	\includegraphics[width=0.8\textwidth]{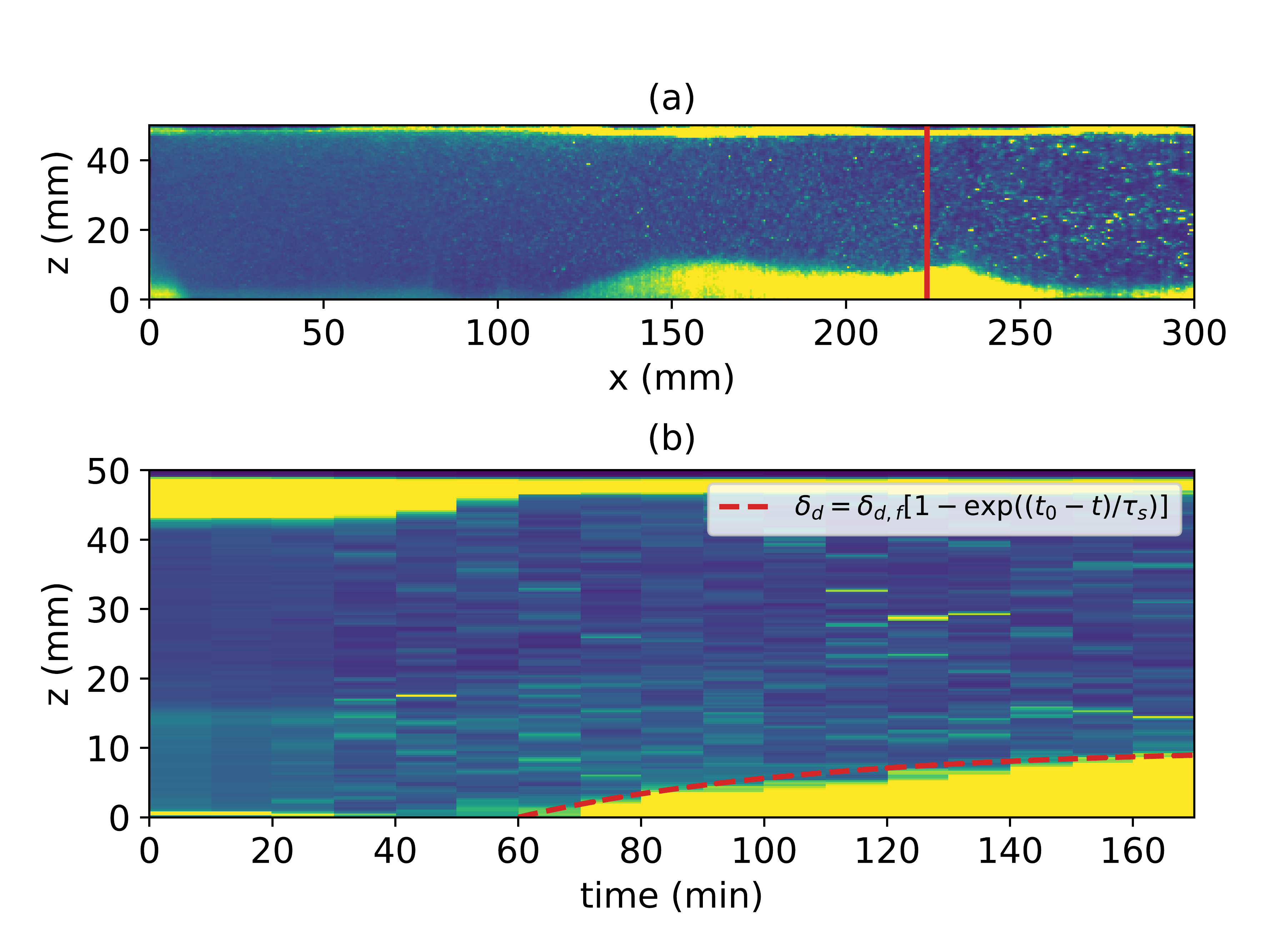}
	\caption{Deposition dynamics in experiment IHB16. (a) Snapshot of the final state of the experiment, showing a stable floating lid and a cumulate. The red vertical line shows the one-pixel slice that is followed as a function of time (b). The cumulate thickness is fitted by an exponential function (dashed line).}
	\label{fig:Cumulate}
\end{figure}
	
	\indent In some experiments we performed, not only is the lid eroded, but also a cumulate forms by sedimentation of the beads from the convective suspension (see Table \ref{tab:Exp}). Indeed the formation of a cumulate is expected when the bulk temperature is high enough for the beads to become negatively buoyant and are able to settle at the base of the tank. The formation of the cumulate always happens in a late stage of the experiments, once the lid has reached its steady state thickness. The evolution of the lid and of the cumulate can then be modeled independently. However, as highlighted by \citeA{Sturtz21}, the criterion for the cumulate formation is still related to the Shields number, that must be subcritical in order to allow beads settle.\\
	\indent To model the formation of the cumulate, we consider a convective suspension with a solid fraction $\phi(t)$ that can only decrease by particle deposition, a valid hypothesis in our case as the floating lid has reached its steady state thickness. We express the mass conservation of beads in the bulk as:
\begin{linenomath*}
	\begin{equation}
		\frac{\rm{d}V_d}{\rm{d}t}=-\frac{\rm{d}V_{\rm{sus}}}{\rm{d}t}, \label{eq:depcons}
	\end{equation}
\end{linenomath*}
where $V_d\sim \phi_{RLP}\delta_d$ is the volume of particles in the cumulate of packing $\phi_{\rm{RLP}}$ and $V_{\rm{sus}}\sim \overline \phi h$ is the volume of particles in suspension with $\overline \phi$ the volume averaged concentration of particles in the bulk.From the expressions of $V_{\rm{sus}}$ and $V_d$ we get:
\begin{linenomath*}
	\begin{equation}
		\frac{\rm{d} \delta_d}{\rm{d} t} \sim -\frac{h}{\phi_{\rm{RLP}}}\, \frac{\rm{d} \overline \phi}{\rm{d} t},
	\end{equation}
\end{linenomath*}
To go further, we use the model of \citeA{Lavorel09} to determine $\phi(t)$. We consider that beads can leave the bulk only if the Shields number is smaller than its critical value, and do so at a settling velocity $v_s$ to form the deposit. As the bulk Shields number is sub-critical, the cumulate is stable and we consider that no re-entrainment occurs from the cumulate formed at the bottom of the tank. All these assumptions put together leads to the evolution equation of the volume averaged concentration of particles $\overline \phi(t)$:
\begin{linenomath*}
	\begin{equation}
		\frac{\rm{d} \overline \phi}{\rm{d} t}=- \frac{v_s}{h} \overline \phi. \label{eq:bulkcons}
	\end{equation}
	\end{linenomath*}
Finally, using (\ref{eq:depcons})  and (\ref{eq:bulkcons}) we get the deposition law:
\begin{linenomath*}
	\begin{equation}
		\frac{\rm{d}\delta_d}{\rm{d}t}=c_d v_s  \frac{ \overline\phi}{\phi_{RLP}}. \label{eq:CumulateEvol}
	\end{equation}
\end{linenomath*}
where $c_d$ depends on bead shape.\\
\indent To further propose a first order solution of (\ref{eq:CumulateEvol}), we consider an initial concentration of beads in suspension $\phi_0$ before deposition, which yields:
\begin{linenomath*}
	\begin{equation}
		\phi=\phi_0\, \frac{h}{h-\delta_d}-\phi_{RLP}\frac{\delta_d}{h}.
	\end{equation}
\end{linenomath*}
As the floating lid is supposed to be at steady state before cumulate starts to form, $\phi_0$ is linked to the difference between the initial and the final lid thicknesses. The solution of (\ref{eq:CumulateEvol}) is an exponential: $\delta_d\sim \rm{exp}(-t /\tau_s)$ where the deposition timescale is:
\begin{linenomath*}
	\begin{equation}
		\tau_s=\frac{1}{c_d} \frac{h}{v_s}.
	\end{equation}
\end{linenomath*}
The problem thus boils down to the determination of the settling velocity $v_s$. To that aim,  we analyse the experiments where a stable deposit formed, as illustrated in Figure \ref{fig:Cumulate}. We choose a vertical slice of the frames at position $x$  (vertical blue line in (a)), and we represent the evolution of this slice over time (b) as a proxy of the evolution of the cumulate. As illustrated in Figure \ref{fig:Cumulate} (b), the cumulate growth can be fitted by the exponential law proposed above. Following several authors \cite{MartinNokes88,MartinNokes89,Lavorel09}, we assume that the settling velocity scales with the Stokes velocity: $v_s\sim \Delta \rho g r^2/\eta_f$.  Using the four experiments with a sufficient cumulate thickness (the 3 other experiments produced a too thin deposit), we obtain $c_d=0.24\pm0.14$. The value of $c_d$ is close to the coefficient $2/9\approx 0.22$ that appears in the expression of the Stokes velocity which means that we do not have to introduce correcting a factor on the viscosity due to convection. The dispersion is due to the inhomogeneity of the cumulate thickness. To reduce the uncertainty, we take 3 values per experiments at different position $x$. The beads radius distribution contributes also to the dispersion (see \citeA{Sturtz21}, supplementary materials). Again, the dispersion can be due to the distribution in bead radius that affects the sedimentation velocity.

\section{Conclusion}

\indent We develop an erosion/deposition model for particles sheared by convection by adapting the Shields' formalism to convective systems. We model the erosion process in granular beds that lie in a thermal boundary layer. The erosion law is a function of the Rayleigh-Roberts number that characterizes convection, and the Shields number that quantifies the ability of beads to be freed from frictional interactions. Within the same framework, we further established and described the subsequent formation of basal cumulates.\\
\indent The model has been validated using lab-scale experiments in a fluid bearing plastic beads and heated from within. The theoretical framework and the scaling laws proposed here can be applied to various geophysical systems, such as magma oceans. Such applications are the subject of a companion paper.

%

\section*{Declaration of interest}
The authors report no conflicts of interest.
		
%
%

\acknowledgments
This paper is part of Cyril Sturtz's PhD thesis (Universit\'e de Paris, Institut de Physique du Globe de Paris). The authors would like to thank \'Eric Lajeunesse and Claude Jaupart for fruitful discussions that have made this article possible. This study contributes to the IdEx Universit\'e de Paris ANR-18-IDEX-0001. This work was supported by the Programme National de Plan\'etologie (PNP) of CNRS/INSU, co-funded by CNES.

%
%

\appendix

%
%

\section{Experimental conditions}
\label{App:Exp}

\begin{table}[h]
	\centering
	\begin{tabular}{c c c c}
	\hline
	Properties 					& Symbol 			&Value 		& Unit\\
	\hline
	Fluid density ($20^{o}\rm{C}$)		& $\rho_{0,f}$ 		& $1192$ 		& $\mathrm{kg\ m^{-3}}$\\   
	Beads density ($20^{o}\rm{C}$)		& $\rho_{0,p}$		& $1187$ 		& $\mathrm{kg\ m^{-3}}$\\
	Fluid thermal expansion 			& $\alpha_{f}$		& $5.5\ 10^{-4}$& $\mathrm{K^{-1}}$\\
	Beads thermal expansion 			& $\alpha_{p}$		& $3.2\ 10^{-4}$& $\mathrm{K^{-1}}$\\
	Fluid viscosity ($20^{o}\rm{C}$)		& $\eta_{f}$		& $0.151$ 	& Pa\ s\\
	Activation energy 				& $E_{a}$			& $41.7$ 		& $\mathrm{kJ\ mol^{-1}}$\\
	Fluid thermal diffusivity			& $\kappa_{f}$		& $9.1\ 10^{-8}$& $\mathrm{m^{2}\ s^{-1}}$\\
	Beads thermal diffusivity (*) & $\kappa_{p}$	& $1.\ 10^{-7}$ 	&$\mathrm{m^{2}\ s^{-1}}$\\
	Fluid thermal conductivity 			& $\lambda_{f}$	& $0.276$ 	&$\mathrm{W\ m^{-1}\ K^{-1} }$\\
	Beads thermal conductivity (*) 		& $\lambda_{p}$	& $0.21$ 		& $\mathrm{W\ m^{-1}\ K^{-1} }$\\
	\hline
	\end{tabular}
	\caption{Main physical properties of the fluid and beads. The activation energy is obtained from the viscosity fit with an Arrhenius law: $\eta(T)=\eta_{f} \exp\left[\dfrac{E_{a}}{R}\left(\dfrac{1}{T}-\dfrac{1}{T_{0}}\right)\right]$, with $T_{0}=20^{o}\rm{C}$. Properties are all measured in the lab, except those marked with (*) which are taken from \cite{HandbookPolymer}. See Supplementary Materials for further information on the way properties measurements have been carried on.}
	\label{tab:ParamsExp}
\end{table}

\section{Continuity of the flux at the base of the floating lid}
\label{App:continuity}
\indent We compare here that the flux out of the convective bulk to the . Following \citeA{Kerr90a,Kerr90b}, the difference between the two fluxes is given by the conservation of heat at the moving interface:
\begin{linenomath*}
	\begin{equation}
		Q_{s,\rm{lid}}(t)=\rho c_p (T_{\rm{bulk}}-T_{\rm{lid}})\, \frac{\rm{d}\delta}{\rm{d} t}+Q_{s,\rm{conv}}(t).
	\end{equation}
	\end{linenomath*}
We take the dimensionless form of this equation by using the characteristic flux $Hh$. Using the scaling law for convection $T_{\rm{bulk}}-T_{\rm{lid}}\sim \Delta T_H Ra_H^{*,-1/4}$ and the evolution equation of the lid thickness given by (\ref{eq:LidEvolution}):
\begin{linenomath*}
	\begin{equation}
		Q_{s,\rm{lid}}^*(t)=\frac{r}{h}(\zeta-\zeta_c)\, Ra_H^{*,1/4}+Q_{s,\rm{conv}}^*(t),
	\end{equation}
\end{linenomath*}
where $Q_{s,i}^*$ stands for dimensionless fluxes. The first term in the right-side of this equation equals $0.01-0.09$ in our experiments, and that is why we assume that $Q_{s,\rm{lid}}(t)\approx Q_{s,\rm{conv}}(t)$.

%

\section{Numerical scheme for the experimental set-up}
\label{App:NS}

\indent To simulate numerically what happens experimentally, we discretized the system of equations over $N_t=6000$ points in time. Thus, the time at the time-step $i$ is written $t_i=i\times \rm{d}t$. Similarly, all parameter $\mathcal{P}$ at $t_i$ is call $\mathcal{P}_i$. At each step, we deduce the evolution of the system by the following numerical scheme, assuming that all parameters are determined at $t_{i-1}$:

\begin{enumerate}

	\item \textbf{Mechanical evolution of the bed :}\\
	if $\zeta_i>\zeta_c$, the bed is eroded, then we calculate the rate of erosion by using the 	following explicit scheme of the equation (\ref{eq:LidEvolution}):
	\begin{linenomath*}
	\begin{equation}
		\delta_{i+1}=\delta_{i}+\frac{\rm{d}\delta}{\rm{d} t} \times \rm{d}t, \label{eq:ErosionNS}
	\end{equation}
	\end{linenomath*}
	with $\rm{d}\delta/\rm{d}t$ given by (\ref{eq:LidEvolution}). For the sake of convenience, we can re-express the (\ref{eq:LidEvolution}) as a function of the lid thickness only. To do so, we re-express $\zeta_{\rm{lid}}$ and $\zeta_c$ as a function of the $\delta$. In this way, we assess that the scaling law determined by \citeA{Sturtz21} for $\zeta_{\rm{lid}}$ remains true in transient state, replacing $Ra_{H}$ by $Ra_{H}^{*}$. Besides, as the bed is thin compared to the tank thickness, the temperature profile in the bed can be assessed to be linear at any time, so that: $T_{\rm{lid}}=T_{s}+Q_{s}(t)\delta(t)/\lambda_{b}$. This assumption is justified by comparing conductive timescale $\tau_{\rm{cond}}\sim\delta^{2}/\kappa_{b}$ to the convective one $\tau_{\rm{conv}}\sim h^{2}/\kappa_{f}\, Ra_{H}^{-1/4}$ \cite{Limare19}. We get $\tau_{\rm{cond}}/\tau_{\rm{conv}}\ll 1$, which corroborates that the lid is at thermal equilibrium at any time. Moreover, $T_{\rm{lid}}=T_{bulk}-C_{T}^*\, \Delta T_{H}^*\, Ra_{H}^{*,-1/4}$ and $\lambda_b$ the floating lid thermal conductivity. Consequentely, (\ref{eq:shields}) becomes:
\begin{linenomath*}
	\begin{equation}
		\zeta_{\rm{lid}}=\frac{\zeta_{s}}{1-\frac{\Delta(\rho_{0}\alpha)}{\Delta \rho_{0}}\frac{\delta(t) Q_{s}(t)}{\lambda_{b}}}.
	\end{equation}
\end{linenomath*}
Similarly, $\zeta_c$ is expressed as a function of $\delta_{\rm{th}}$, the thickness of the bed at steady state:
\begin{linenomath*}
	\begin{equation}
		\zeta_{c}=\frac{\zeta_{s}}{1-\frac{\Delta(\rho_{0}\alpha)}{\Delta \rho_{0}}\frac{\delta_{\rm{th}}Hh}{\lambda_{b}}}.
	\end{equation}
\end{linenomath*}
The erosion of the bed (\ref{eq:LidEvolution}) is therefore expressed as a function of the thickness of the bed at steady state $\delta_{\rm{th}}$, which can be calculated from the thermal steady state of each experiments by combining (\ref{eq:DTBL}) and assuming that the heat flux through the floating lid is given by $Hh=\lambda_b (T_{\rm{lid}}-T_s)/\delta_{\rm{th}}$, \cite{Sturtz21}:
\begin{equation}
	\frac{\delta_{\rm{th}}}{h}=\frac{\lambda}{\lambda_f}\, \left(\frac{T_{\rm{bulk}}-T_s}{\Delta T_H}-C_T\, Ra_H^{-1/4} \right),
\end{equation}
where $T_{\rm{bulk}}$ is the volume average temperature of the convective bulk at steady state.

If $\zeta_i<\zeta_c$, deposition is possible.\\
	As the lid does not absorbe microwave, we calculate the real rate of internal heating $H_{i+1}=H_0\times h/\delta_{i+1}$.
	
	\item \textbf{Basal temperature of the floating lid:}\\
	we use an explicit scheme of (\ref{eq:ConsNRJ}):
	\begin{linenomath*}
	\begin{equation}
		T_{\rm{bulk,i+1}}=T_{\rm{bulk},i}+\frac{1}{\rho_{\rm{f}}c_{p,\rm{f}}} \left(H_i- \frac{Q_{s,\rm{conv},i}}{h-\delta_i}\right)\times \rm{d}t.
	\end{equation}
	\end{linenomath*}
	
	\item \textbf{Thermal state of the bed :}
	\begin{itemize}
		\item \textbf{If the lid thickness does not change:} we compute equations	(\ref{eq:TS_b1})-(\ref{eq:TS_b4}) using a 1 dimension implicit finite difference scheme. The lid is discretized in $N_z=500$ points, and the position in the lid is called $z_j=j\times \rm{d}z$. The temperature field in the bed at time-step $i$ is represented by the $N_z$-long vector $\mathcal{T}_i$. The temperature at position $z_j$ and time $t_i$ is noted $T_{b,i}^j$. Thus, the temperature in the floating lid at the time-step $t_{i+1}$ can be calculated as follow :
	\begin{linenomath*}
	\begin{equation}
		\mathcal{T}_{i+1}=\mathcal{A}^{-1}_i\cdot \mathcal{T}_i,
	\end{equation}
	\end{linenomath*}	
	where
	\begin{linenomath*}
	
	\begin{equation}
		\mathcal{A}_i=
		\begin{pmatrix}
		\epsilon		& 0 		& \cdots	& 		&  		& 0 \\
		\hline
		-\epsilon 	& C		& -\epsilon & 0 		& \cdots	& 0 \\
		\vdots 	&\ddots	& \ddots	&\ddots	&  		&  \vdots\\
		\vdots 	& 		& \ddots	&\ddots	&\ddots 	&  \vdots\\
		0	 	& \cdots	& 	0 	& -\epsilon &  	C	& -\epsilon\\ 
		\hline
		0	 	& \cdots	& 	0 	& -\epsilon &  	-4\epsilon 	&  -3\epsilon\\
		\end{pmatrix}
		\hspace{0.5cm}
		\rm{and}
		\hspace{0.5cm}
	\mathcal{T}_i=
		\begin{pmatrix}
			T_s\times \epsilon  \\
			\hline
			T_{b,i}^2\\
			\vdots\\
			T_{b,i}^j\\
			\vdots\\
			T_{b,i}^{N_z-1}\\
			\hline
			2 \rm{d} z\, \frac{Q_{s,\rm{conv},i}}{\lambda_b}\, \epsilon\\
		\end{pmatrix}
		\end{equation}
		
	\end{linenomath*} 
	
		where $\epsilon=\kappa_b \rm{d}t/\rm{d}z$, $C=1+2\epsilon$ and $T_{b,i}^j$ is the temperature field in the floating lid at the previous time-step. We included the boundary conditions in the matrix. The Dirichlet condition at $z_j=0$ is a fixed surface temperature : $T_{b,i}^{j=0}=T_s$. The Neumann condition at the base of the floating lid is due to the fixed flux imposed by the convection, and is calculated by using an order 2 approximation for the temperature gradient at $z_{j=N_z}=\delta_{i+1}$ :

	\begin{linenomath*}
	\begin{equation}
		\left[ \frac{\rm{d}T_b}{\rm{d}z}\right]_{N_z}\approx \frac{T_{b,i}^{Nz-2}-4T_{b,i}^{Nz-1}+3T_{b,i}^{Nz}}{\rm{d}z}=\frac{Q_{s,\rm{conv},i}}{\lambda_f}.
	\end{equation}
	\end{linenomath*}
	\end{itemize}
	
	\item \textbf{If the lid thickness has evolved:} we adapted the grid discretizing the lid in order to always be regular and have $N_z=500$ points. \\
	\indent If the lid is eroded, we use only the temperature field at $t_i$. Necessarily, the thickness $\delta_{i+1}$ is bracketed between $z_{j_{max}}<\delta_{i+1}$ and $z_{j_{max}}>\delta_{i+1}$. Thus, we can estimated the temperature at $\delta_{i+1}$ by an interpolation of the temperature field between $z_{j_{max}}$ and $z_{j_{max}}$.\\
	\indent If the lid thickens, we add one row at $T^i$, corresponding to the position $\delta_{i+1}$ and the corresponding temperature is estimated by an extrapolation of $T_{b,Nz}^i$ using the basal heat flux. Then, this field is resample on a regular grid of size $N_z$.\\
	\indent Once the lid evolves thermally, we deduce the basal temperature of the lid $T_{lid,i+1}=T_{b,i+1}^{N_z}$. If the lid is totally eroded, $T_{lid,i+1}=T_s$.
	
	\item \textbf{Calculation of the convective heat flux :}
	we use the scaling law (\ref{eq:ScalQs}) :
	\begin{linenomath*}
	\begin{equation}
		Q_{s,\rm{conv},i+1}=\frac{\lambda_f}{C^{*,4/3}_T}\, \left(\frac{\alpha_f \rho_f g}{\kappa_f \eta_f}\right)^{1/3}\left(T_{\rm{bulk},i+1}-T_{\rm{lid},i+1}\right)^{4/3}
	\end{equation}
	\end{linenomath*}
	
	\item \textbf{Calculation of all parameters :}\\
	We calculate all dimensionless numbers that can be important : the Rayleigh-Roberts number $Ra_{H,i+1}^*$, and the Shields number $\zeta_{i+1}$.

\end{enumerate}

\bibliography{biblio}

\end{document}